\documentclass[opre]{informs3nothing}  

\DoubleSpacedXI 

\usepackage{latexsym, bbm, enumerate, amssymb, amsmath, libertine,tikz}
\usepackage{gensymb}
\usepackage{booktabs}
\usepackage{float}
\usepackage[utf8]{inputenc}
\usepackage[T1]{fontenc}
\usepackage{subcaption}

\usepackage{natbib}
\usepackage{hyperref}
\bibpunct[, ]{(}{)}{,}{a}{}{,}%
\def\bibfont{\small}%
\graphicspath{{./figures/}}

\TheoremsNumberedThrough     
\ECRepeatTheorems

\EquationsNumberedThrough    

\newcommand{\E}{\mathbb{E}}

\DeclareMathOperator{\Var}{Var}
\DeclareMathOperator{\Cov}{Cov}

\newcommand{\paranth}[1]{\left(#1\right)}
\newcommand{\bracket}[1]{\left[#1\right]}
\newcommand{\curly}[1]{\left\{#1\right\}}

\TITLE{Pairs Trading Using a Novel Graphical Matching  Approach}
\RUNTITLE{Pairs Trading Using a Novel Graphical Matching  Approach}

\RUNAUTHOR{Qureshi and Zaman}
\ARTICLEAUTHORS{%
	\AUTHOR{Khizar Qureshi}
	\AFF{%
		Massachusetts Institute of Technology, 
		77 Massachusetts Ave.,
		Cambridge, MA~  02139, 
		\EMAIL{kqureshi@mit.edu}
	}
	\AUTHOR{Tauhid Zaman}
	\AFF{%
		Yale School of Management, Yale University,  
		165 Whitney Ave Ave., 
		New Haven, CT~ 06511,  
		\EMAIL{tauhid.zaman@yale.edu}
	}
} 
\ABSTRACT{
    Pairs trading, a strategy that capitalizes on price movements of asset pairs driven by similar factors, has gained significant popularity among traders. Common practice involves selecting highly cointegrated pairs to form a portfolio, which often leads to the inclusion of multiple pairs sharing common assets. This approach, while intuitive, inadvertently elevates portfolio variance and diminishes risk-adjusted returns by concentrating on a small number of highly cointegrated assets. Our study introduces an innovative pair selection method employing graphical matchings designed to tackle this challenge.  We model all assets and their cointegration levels with a weighted graph, where edges signify pairs and their weights indicate the extent of cointegration.  A portfolio of pairs is a subgraph of this graph.  We construct a portfolio which is a maximum weighted matching of this graph to select pairs which have strong cointegration while simultaneously ensuring that there are no shared assets within any pair of pairs.  This approach ensures each asset is included in just one pair, leading to a significantly lower variance in the matching-based portfolio compared to a baseline approach that selects pairs purely based on cointegration.   Theoretical analysis and empirical testing using data from the S\&P 500 between 2017 and 2023, affirm the efficacy of our method. Notably, our matching-based strategy showcases a marked improvement in risk-adjusted performance, evidenced by a gross Sharpe ratio of 1.23, a significant enhancement over the baseline value of 0.48 and market value of 0.59. Additionally, our approach demonstrates reduced trading costs attributable to lower turnover, alongside minimized single asset risk due to a more diversified asset base.

}

\begin{document}
\maketitle
\section{Introduction}
 Pairs trading is an investment strategy based on the co-movement of prices of pairs of assets. If the two prices, typically driven by common statistical factors, diverge, a long-short position in the two assets can be used to profit from the re-convergence. Pairs are selected based on statistical price similarity or fundamental similarity between firms, and then monitored for price divergence from pre-determined thresholds. Based on the distance from the threshold, traders buy one of the assets, and sell the other, in order to establish a long-short position until the spread between the asset prices returns to its expected value. A comprehensive review of pairs trading can be found in \cite{vidyamurthy}.

Literature has maintained a focus on identifying single pairs and monetizing on a on a daily, monthly, or annual basis. However, there is a larger issue in a portfolio of pairs that has not yet been addressed extensively. A pairs trading investment strategy involves trading on not just a single pair, but rather a large number number of pairs to maximize the expected number of trading opportunities. At any given time, some pairs will have convergence opportunities while others will not. Although a large number of pairs will increase the expected number of opportunities, it will do so with two costs: (i) transaction costs and (ii) covariance. The former is straightforward -- there are two-way execution costs and slippage associated with each investment into a pair. The latter, which we will formalize later, is due to the fact that distinct pairs may share assets. Suppose for example, that there are two pairs of stocks (1,2) and (1,3). Although both are distinct pairs, they share in common stock 3, and consequently, the spreads for the pairs may have a positive covariance. In turn, this covariance will increase the risk of the overall portfolio by increasing its variance. With a very large number of pairs, we find that this has a significant impact on the risk-adjusted performance and draw-down of the investment strategy.

\subsection{Our Contribution}

One popular method of pairs selection is through cointegration, which involves testing whether the linear combination of two time series is stationary \citep{walter}. In practice however, traders may neglect the covariance between asset pairs simply because any two given pairs are for the most part ineligible for a pairs investment strategy. However, conditional on the set of cointegrated pairs, we find that the covariance term makes a large impact downstream on the strategy performance. In order to reduce the covariance term, we propose techniques from graph theory such as matchings, to determine subsets of pairs with no common assets. We  formally prove that the covariance from shared stocks in typical pair portfolio construction methods is substantial, and detrimental to the risk-adjusted performance. We then propose and implement a matching-based pair selection method to prevent the prior effect. To the best of our knowledge, this is the first instance in the literature in which the impact of covariance/variance in pairs trading portfolios is identified and in which a graphical matching-based approach is used to select pairs.  

\subsection{Previous Work}

Most pairs trading literature is oriented around price-based signals. In the work of \cite{gatev}, the authors select pairs by minimizing the distance between normalized stock prices, and conclude that gross profitability is linked to common factors between stocks. However, the authors also find that profits sometimes exceed transaction cost estimates. An extension by \cite{elliott} uses the expectation maximization (EM) algorithm to update a Gaussian Markov chain model for the pair spread. 
Other authors use timing, rather than critical price levels to identify opportunities. For example, \cite{jurek} shows opportunity in a learning approach, by using intertemporal hedging and methods of capturing mispricing to show there are critical levels in which profitability exists in the trade. Similarly, \cite{liu} show that conditioning on liquidation (or lack-thereof) on one asset of the pair can increase profitability through expected convergence. In the work of \cite{avellaneda}, the authors use PCA-based trading strategies to show efficacy in ETF pairs trading  as well as an enhancement using trading volume features. 

Another cohort of authors have explored pairs trading through the use of copula, which is a multivariate cumulative distribution function such that the marginal probability of each variable is uniform in a $[0,1]$ measure, and are used to describe dependence between random variables. For example, \cite{liew} constructs conditional probabilities using the derivative of the copula of various distributions, and applies it as a trigger for traditional distance and and cointegration based approaches. On the other hand, \cite{botha} uses bivariate copula to identify confidence intervals for two highly correlated stocks' residual spread. For both approaches, although the copula is effective against the benchmark, the benefit is marginal, and consumed by transaction costs.  In terms of initial trading universe, most prominent papers have used one of the various indices in the S\&P 500. For example, \cite{huck2} uses the S\&P 500 to show that cointegration methods generally dominate distance-based methods, and similarly, \cite{huck} uses the S\&P 100 to establish a ranking framework for pair selection. Pairs trading is not limited to equities, and is applicable to other asset classes. For example, \cite{cummins} using various statistical learning algorithms to show profitability in traditional spread-trading on the crude oil and refinery product markets.

\section{Pairs Trading Background}

The traditional pairs trading framework has three main components: (i)  identification, (ii) selection based on cointegration testing, and (iii) trading strategy. First, the identification of potential pairs involves selecting tuples of assets whose prices may move together in short time intervals. In itself, this is a forecast, and can involve (i) identifying companies within a particular sector such as technology or health care, (ii) identifying companies impacted by the same risk factors, or even (iii) identifying companies who share supply chain commitment. 

The second component, selection based on cointegration, involves testing for a statistical property between the log prices of the two assets. The order of integration, $I(d)$ of a time series is the minimum number of differences that must be carried out to obtain a covariance-stationary time series. Specifically, a time series is integrated of order $d$ if $(1-L)^d X_t$ is a stationary process for some lag operator $L$. For example, if $d=0$, and the time series is of order zero, then $(1-L)^0 X_t = X_t$. With an understanding of order of integration, we can now define cointegration. For a pair of time series to be cointegrated, there are two requirements: (i) the log prices must be integrated with some order $d$ (ii) the linear combination of the two time series must be integrated of order less than $d$. If these two conditions hold, the aforementioned linear combination, or spread is cointegrated. The most common test for cointegration is the Engle-Granger Two-Step Method, but there are several others including the Johansen test, Phillips-Ouliaris test, and Bayesian inference. A comprehensive overview of these methods can be found in \cite{walter}. 

The third component, trading, involves tracking the spread between two or more assets and simultaneously buying and selling it at outlier levels until it returns to equilibrium. The expectation of the spread is measurable over time, and can be used to represent the long-term equilibrium. Consequently, the distance from the equilibrium value can behave as a forecast if we believe there will be a convergence or return to the equilibrium value in the future. To act on the forecast, we can buy one of the assets and sell the other, effectively buying or selling the spread if it is below or above the equilibrium value. The position is profitable only if the spread reverts to its equilibrium level, so traders selectively enter and exit these spreads when they deviate sufficiently.

\section{Theoretical Analysis of Pairs Trading}\label{sec:theory}
We will consider a portfolio consisting of both cointegrated and non-cointegrated pairs of stocks. To assess the portfolio's performance and the efficacy of our strategy for selecting pairs, it's crucial to understand the mean and variance of the returns of these pairs, along with the covariance among the returns of different pairs. In this section, we introduce simple models to represent the behavior of pairs in both cointegrated and non-cointegrated contexts. Following this, we will derive  formulas for analyzing the returns. These formulas will be instrumental in evaluating the theoretical performance of various portfolio compositions, highlighting the benefits of our matching-based pair selection strategy compared to baseline methods.  The proofs of the theoretical results in this section are provided in the Appendix.

\subsection{Cointegrated Pairs}
We will trade a pair involving stocks 1 and 2 which are cointegrated.  The prices of the stocks at discrete time $t$ are given by $p_{1t}$ and $p_{2t}$.  We assume the log-price for stock 1 follows a random walk.  That is, 
\begin{align}
    \log(p_{1,t+1}) & = \log(p_{1t}) + \mu_1 +  \delta_{1,t+1} \label{eq:cointegrated_price}
\end{align}
where $\mu_1$ is a constant mean term, and the noise terms  $\delta_{1t}$ are independent and identically distributed (i.i.d.) normal random variables with mean zero and standard deviation $\sigma_1$.  Because stocks 1 and 2 are cointegrated,  their prices are related by

\begin{align}
    \epsilon_t & =\log(p_{2t}) -\log(p_{1t})  \label{eq:cointegrated_spread} 
\end{align}
where $\epsilon_t$ are i.i.d. normal random variables with mean zero and standard deviation $\sigma$. We refer to  $\epsilon_t$ as the \emph{spread} at time $t$.  Trading a pair of stocks is based upon the value of the observed spread.  To trade the pair, we choose a threshold $k\sigma$ for some $k>0$.  The trading signal is given by 
        \begin{align}
            S_t =  \mathbbm{1}\curly{\epsilon_t \leq -k\sigma} -   \mathbbm{1}\curly{\epsilon_t \geq k\sigma}\label{eq:signal}.
        \end{align}
This signal is negative if the spread is too low, meaning the price of stock 2 is too low relative to stock 1.  In this case we want to sell stock 1 and buy stock 2.  If the signal is positive then the spread is too high, meaning the price of stock 2 is too high relative to stock 1.  In this case we want to buy stock 1 and sell stock 2. The return of stock $i$  at time $t+1$ is
        $$r_{i,t+1} = \frac{p_{i,t+1}}{{p_{it}}} -1$$
and the return of the pair at time $t+1$ is 
\begin{align}
    r_{t+1} &= S_t(r_{2,t+1} - r_{1,t+1}). \label{eq:cointegrated_return}
\end{align}
We have the following result for the mean and variance of the cointegrated pair returns.
\begin{theorem}\label{thm:cointegrated_mean_var}
    Consider a cointegrated pair of stocks 1 and 2 with prices given by equations \eqref{eq:cointegrated_price}
    and \eqref{eq:cointegrated_spread} traded using the signal given by equation \eqref{eq:signal}, with returns given by equation \eqref{eq:cointegrated_return}.    
    Let $Z$ be a standard normal random variable.  Then the pair return at time $t$ has mean
    \begin{align}
         \mathbf E[r_{t}] & = e^{\mu_1+\sigma_1^2/2+\sigma^2}
                               \mathbf P( k-\sigma \leq Z \leq k+\sigma)   \label{eq:returns_cointegrated_mean}
    \end{align}
    and variance 
    \begin{align}
 \text{Var}(r_{t+t})  =& e^{2\mu_1+2\sigma_1^2+4\sigma^2} \paranth{\mathbf P(Z\leq -k+2\sigma) + \mathbf P(Z\geq k+2\sigma)}\nonumber\\
                    &-2e^{2\mu_1+2\sigma_1^2+\sigma^2}\paranth{\mathbf P(Z\leq -k+\sigma) + \mathbf P(Z\geq k+\sigma)}\nonumber\\ 
                    &+2e^{2\mu_1+2\sigma_1^2}\mathbf P(Z\geq k)- \mathbf E^2[r_t].\label{eq:returns_cointegrated_var}
\end{align}
\end{theorem}
From this result, it can be shown that the mean return of a cointegrated pair is positive, which is why they provide a profitable trading opportunity.  In fact, the mean is positive even if the drift $\mu_1$ is negative.  This means that a pair of stocks whose prices have a negative drift can still be traded as a pair for a profit.  Furthermore it can be shown that the mean return of the pair increases for larger standard deviations $\sigma_1$ and $\sigma$, which indicates that more volatile pairs are more profitable.


We next consider stocks 1, 2, 3 and pairs $a$ = (stock 1, stock 2) and $b$ = (stock 1, stock 3).  We assume each pair is cointegrated.  Then the spreads of the two pairs can be written as
\begin{align}
    \epsilon_{at} & =\log(p_{2t})- \log(p_{1t}) \label{eq:cointegrated_spread_a} \\
    \epsilon_{bt} & =\log(p_{3t}) -\log(p_{1t})  \label{eq:cointegrated_spread_b}.
\end{align}
The spreads $\epsilon_{at}$ and $\epsilon_{bt}$ have zero mean and standard deviations $\sigma_a$ and $\sigma_b$, respectively.  The signal for each pair is 
 \begin{align}
            S_{at} =  \mathbbm{1}\curly{\epsilon_{at} \leq -k\sigma_a} -   \mathbbm{1}\curly{\epsilon_{at} \geq k\sigma_a} \label{eq:signal_a}\\
            S_{bt} =  \mathbbm{1}\curly{\epsilon_{bt} \leq -k\sigma_b} -   \mathbbm{1}\curly{\epsilon_{bt} \geq k\sigma_b} \label{eq:signal_b}.
        \end{align}
The returns of the spreads are
\begin{align}
    r_{a,t+1} &= S_{at}(r_{2,t+1} - r_{1,t+1}) \label{eq:cointegrated_return_a}\\
    r_{b,t+1} &= S_{bt}(r_{3,t+1} - r_{1,t+1}) \label{eq:cointegrated_return_b}
\end{align}
The sharing of stock 1 between the two pairs creates a positive covariance between the returns.   
The following result gives an expression for this covariance.
\begin{theorem}\label{thm:cov_cointegrated}
 Consider stocks 1, 2, and 3 with prices given by equations \eqref{eq:cointegrated_price},
    \eqref{eq:cointegrated_spread_a}, and \eqref{eq:cointegrated_spread_b}.  Define the pairs $a$= (stock 1, stock 2) and $b$ = (stock 1, stock 3).  Assume the pairs are traded using the signals given by equations \eqref{eq:signal_a} and \eqref{eq:signal_b}, with returns given by equations \eqref{eq:cointegrated_return_a} and \eqref{eq:cointegrated_return_b}.     Then the covariance of the returns of the pairs at time $t$ is
    \begin{align}
     \text{Cov}(r^a_{t},r^b_{t}) & = \paranth{e^{\sigma_1^2}-1}\mathbf E[r_{at}]\mathbf E[r_{bt}]\label{eq:cointegrated_cov}.
\end{align}    
\end{theorem}
Because the mean returns of cointegrated pairs are positive, Theorem \ref{thm:cov_cointegrated} shows that the returns of two pairs which share a common stock will have a positive covariance.

\subsection{Non-Cointegrated Pairs}
We next study the statistics of the returns of pairs which are not cointegrated.  In this case we assume the log-price for each stock $i$  follows an independent random walk.  That is, 
\begin{align}
    \log(p_{i,t+1}) & = \log(p_{it}) + \mu_i +  \delta_{i,t+1} \label{eq:noncointegrated_price}
\end{align}
where $\mu_i$ is a constant mean term, and the noise terms  $\delta_{it}$ are i.i.d. normal random variables with mean zero and standard deviation $\sigma_i$.  Unlike with cointegrated stocks, here each stock $i$ follows a log-normal model with independent noise terms and there is no cointegration relationship such as that given by equation \eqref{eq:cointegrated_spread}. Assume the pair is (stock 1, stock 2).  We define the spread of the pair at time $t$ as

\begin{align}
    \epsilon_{t} & = \log(p_{2t}) -\log(p_{1t}) - (\mu_2-\mu_1)t. \label{eq:noncointegrated_spread}
\end{align}
By subtracting the mean price difference $(\mu_2-\mu_1)t$, the spread becomes a zero mean normal random variable with variance $\sigma^2 = t(\sigma_1^2 + \sigma_2^2)$.  The trading signal is given by equation \eqref{eq:signal} and the pair return is given by equation \eqref{eq:cointegrated_return}.  We have the following results for the mean and variance of the returns of a pair that is not cointegrated.

\begin{theorem}\label{thm:noncointegrated_mean_var}
   Consider a non-cointegrated pair of stocks 1 and 2 with prices given by equations \eqref{eq:noncointegrated_price}, spread given by equation \eqref{eq:noncointegrated_spread},
   traded using the signal given by equation \eqref{eq:signal}, with returns given by equation \eqref{eq:cointegrated_return}.    
    Let $Z$ be a standard normal random variable.  Then the pair return at time $t$ has mean
    \begin{align*}
        \mathbf E[r_{t}]=0
    \end{align*}
    and variance
\begin{align*}
    \text{Var}(r_{t}) & = 2\mathbf P(Z\geq k)\paranth{e^{2\mu_1+2\sigma^2_1} + e^{2\mu_2+2\sigma^2_2}
                                    -2 e^{\mu_1+\mu_2+\sigma^2_1/2+\sigma^2_2/2}}.
\end{align*}
\end{theorem}
From this we see that trading a non-cointegrated pair gives zero profit on average.


We next consider stocks 1, 2, 3 and pairs $a$ = (stock 1, stock 2) and $b$ = (stock 1, stock 3).  Call these pairs $a$ and $b$, and assume each pair is not cointegrated.  Then the spreads of the two pairs can be written as
\begin{align}
    \epsilon_{at} & = \log(p_{2t}) -\log(p_{1t}) - (\mu_2-\mu_1)t. \label{eq:noncointegrated_spread_a}\\
    \epsilon_{bt} & = \log(p_{3t}) -\log(p_{1t}) - (\mu_3-\mu_1)t. \label{eq:noncointegrated_spread_b}
\end{align}
The spreads $\epsilon_{at}$ and $\epsilon_{bt}$ have zero mean and variance $\sigma_a^2 = t(\sigma_1^2 + \sigma_2^2)$ and $\sigma_b^2 = t(\sigma_1^2 + \sigma_3^2)$, respectively. The signal and returns for the pairs follow the same expressions as for the cointegrated pairs.  The following result gives an expression for the covariance of the returns of these two non-cointegrated pairs.

\begin{theorem}\label{thm:cov_notcointegrated}
 Consider stocks 1, 2, and 3 with prices given by equations \eqref{eq:noncointegrated_price}.  Define the pairs $a$= (stock 1, stock 2) and $b$ = (stock 1, stock 3).  Let the spreads for each pair be given by equations
 \eqref{eq:noncointegrated_spread_a} and  \eqref{eq:noncointegrated_spread_b}.
 Assume the pairs are traded using the signals given by equations \eqref{eq:signal_a} and \eqref{eq:signal_b}, with returns given by equations \eqref{eq:cointegrated_return_a} and \eqref{eq:cointegrated_return_b}.    Define $Z_a$ and $Z_b$ as zero mean jointly normal random variables with covariance matrix 
\begin{align*}
    \Sigma & = \begin{bmatrix}
                1 & \frac{\sigma_1^2}{\sqrt{(\sigma_1^2 + \sigma_2^2)(\sigma_1^2+\sigma_3^2)}} \\
                 \frac{\sigma_1^2}{\sqrt{(\sigma_1^2 + \sigma_2^2)(\sigma_1^2+\sigma_3^2)}} & 1 \\
               \end{bmatrix}.
\end{align*}
 Then the covariance of the returns of pairs $a$ and $b$ at time $t$ is
\begin{align*}
    \Cov(r^a_{t}, r^b_{t}) =& 2\paranth{\mathbf P\paranth{Z_a\geq k \bigcap Z_b\geq k }
                          -\mathbf P\paranth{Z_a\geq k \bigcap Z_b\leq -k }}\\
                          &\times\paranth{e^{\mu_2+\mu_3 + \sigma_2^2/2 + \sigma_3^2/2}  
                                         -e^{\mu_1+\mu_3 + \sigma_1^2/2 + \sigma_3^2/2}
                                         -e^{\mu_1+\mu_2 + \sigma_1^2/2 + \sigma_2^2/2}
                                         +e^{2\mu_1 + 2\sigma_1^2}}.
\end{align*}    
\end{theorem}

\subsection{Theoretical Portfolio Performance}\label{ssec:sharpe}
With the results for the mean, variance, and covariance of various types of pairs, we can now compare the performance of different pair portfolios.  In practice, a cointegration test is applied to a universe of pairs.  This test produces a p-value, indicating whether or not the pair is cointegrated.  Then all pairs which have a p-value lower than a fixed threshold are included in the portfolio.  We refer to this as the \textit{baseline} portfolio.  This portfolio may contain pairs which are cointegrated, but it may also contain pairs which are not, but have p-values below the threshold by chance.  The pairs universe is large, so many such false positives can occur.  For instance, for the S\&P 500, there are 500 stocks, which gives 125,000 possible pairs.  If all pairs were not cointegrated, a portfolio built from a p-value threshold of 0.01 with no multiple-hypothesis testing correction would contain approximately 1,250 non-cointegrated pairs.   Our other portfolio is referred to as a \textit{matching} portfolio.  This portfolio selects the pairs using a matching algorithm so that the final pairs in the portfolio do not share any common stocks.  This portfolio can contain a maximum of 250 pairs for the S\&P 500.  

For the baseline portfolio we assume 250 pairs are selected for trading based on their p-values. Each pair is traded using a threshold set by $k=2$.  For each portfolio, we assume a unit amount is invested  in each pair.  To further simplify our analysis, we will assume all daily log-normal price changes have the same mean $\mu_1$ and noise standard deviation $\sigma_1$, and all cointegrated spreads have the same standard deviation $\sigma$.

For each portfolio, we assume it has $n_1=1$ cointegrated pair and $n_2$ non-cointegrated pairs.  We also assume that there are $m_1$ pairs of cointegrated pairs that share at least one stock and  $m_2$ pairs of non-cointegrated pairs that share at least one stock.  We do not assume any stocks are shared between cointegrated and non-cointegrated pairs.

The mean return of a single cointegrated pair at time $t$ is denoted $\mu_c$ and the portfolio return at time $t$ is denoted $R_t$. The mean return of the portfolio is given by $\E[R_t] = n_1\mu_c$.  We see that the only contributors to the mean return are the $n_1$ cointegrated pairs.  To compute the portfolio variance, we define four different terms. Let $\nu_1$ be the variance of a single cointegrated pair, $\nu_2$ be the variance of a single non-cointegrated pair, $\kappa_1$ be the covariance of a pair of cointegrated pairs that share a common stock, and $\kappa_2$ be the covariance of a pair of non-cointegrated pairs that share a common stock. The portfolio variance is given by $
    \Var(R_{t})  = n_1\nu_1 +n_2\nu_2 + 2m_1\kappa_1 +2 m_2\kappa_2$.
We assume the covariance is zero between any pairs which do not share a common stock.

To compare the performance of these portfolios, we estimate values for the different parameters using data and compute all moments using Theorems \ref{thm:cointegrated_mean_var}, \ref{thm:cov_cointegrated}, \ref{thm:noncointegrated_mean_var}, and \ref{thm:cov_notcointegrated}.  We estimated the mean and variance for the daily change in log-prices of all stocks on the S\&P 500 using price data from 2010 to 2023.  We took the medians of these values across all stocks to obtain $\mu_1$ and $\sigma^2_1$.  To estimate the spread variance for cointegrated pairs, we looked at all pairs in the S\&P 500 from January 2021 to  August 2023.  We did a linear regression of the log prices of each pair and measured the variance of the regression residual.  We then performed a single-lag augmented Dickey-Fuller test on the residuals to obtain their p-value for being cointegrated.  This is was repeated once for every month.  Details on this procedure are found in Section \ref{ssec:trading}.  We kept the variance for the 250 pairs  with the lowest p-value each month. We then took the median value of this set of variances across all stocks and months in our dataset to obtain $\sigma^2$.  The values for $\mu_1$, $\sigma_1$, and $\sigma$ are shown in Table \ref{table:params}.

The matching portfolio has $m_1 = m_2 = 0$ and the baseline portfolio has $m_1=0$.  To obtain $m_2$ for the baseline portfolio, we look at the monthly portfolios of the 250 pairs with the smallest p-values in the S\&P 500 from January 2021 to  August 2023.  We calculate the number of pairs of pairs each month that share a common stock, and take the median of these values across all months to obtain $m_2=1,748$.  To get a sense for the magnitude of this value, we consider the case where pairs are chosen at random.  In this case, a portfolio with $250$ pairs can be viewed as a random graph on 500 nodes with $250$ edges. We can approximate this graph as an Erdos-Renyi graph $G(n,p)$ with $n=500$ and $p =0.002$ (we set $p$ equal to the number of edges divided by the total number of possible edges on 500 nodes).  Every edge on this graph is a pair in the portfolio, and every two-hop path on this graph corresponds to a pair or pairs that share a common stock (the node in the middle of the path is the shared stock).  Therefore, $m_2$ is equal to the number of such two-hop paths.  This can easily be estimated using results from random graph theory \citep{bollobas2001random}.  Every triplet of nodes with at least two edges contributes three such paths, and each of these triplets exists with probability $p^2$.    To obtain the mean number of such paths we multiply $3p^2$ by the number of triplets, which gives $3p^2n(n-1)(n-2)/6$.  Substituting $p=0.002$ and $n=500$ into this expression, we get that the mean number of pairs of pairs sharing a common stock is 249.5.  This is nearly seven times smaller than the value seen in the data.  This suggests two things.  First, pairs with low p-values tend to aggregate around a few stocks.  Second, the contribution to the baseline portfolio's variance from pairs which share a pair is much larger than predicted by random graph theory.  Because of these reason, we expect that the matching portfolio should have a much better risk adjusted performance than the baseline portfolio.

Table \ref{table:params} shows the theoretical annualized Sharpe ratios of the two portfolios. We see that the Sharpe ratio of the matching  portfolio is much  higher than the baseline.  Most of this difference comes from the covariance of the pairs with shared stocks, which represents 50\% of the variance in the baseline portfolio.  This covariance term does not exist in the matching portfolio, producing a much lower portfolio variance.  This implies that matching-based pair selection significantly reduces the variance of the pair portfolio relative to the baseline by avoiding pairs sharing stocks.  We will see in Section \ref{sec:results} that the performance improvement of the matching portfolio  predicted by our theory is close to what is seen in trading simulations.

\begin{table}[h]
    \centering
    \begin{tabular}{|c|c|c|c|}
    \hline
    Parameter Description& Parameter & Baseline  & Matching  \\
               &           & Portfolio  & Portfolio\\\hline
    Daily log-price change mean & $\mu_1$ & 0.0005 & 0.0005 \\\hline
    Daily log-price change  std. dev. & $\sigma_1$ & 0.0180 & 0.0180 \\\hline
    Spread std. dev. & $\sigma$ & 0.0711  & 0.0711\\\hline
    Trading threshold & k & 2 & 2\\\hline
    Number of cointegrated pairs & $n_1$ & 1 & 1\\\hline
    Number of non-cointegrated pairs & $n_2$ & 249 & 249\\\hline
    Number of pairs of cointegrated pairs  & $m_1$ & 0 & 0\\
     sharing a common stock & & & \\\hline
    Number of pairs of non cointegrated pairs  & $m_2$ & 1,748 & 0\\
     sharing a common stock & & & \\\hline
     Annualized Sharpe ratio & & 0.50 & 1.18 \\\hline
    \end{tabular}
    \caption{Parameter values used to evaluate performance of baseline and matching pairs portfolios.}
    \label{table:params}
\end{table}

\section{Constructing and Trading a Matching Pairs Portfolio}
We have seen with simple analysis that there is substantial improvement in risk-adjusted portfolio performance  in using matching-based pair selection versus a baseline approach based only on p-values of cointegration tests. We now show how to construct a pairs portfolio using matchings and trade the resulting pairs.  Our approach first selects the pairs for the portfolio at the beginning of a trading period, and then trades these pairs for the duration of the period.  

\subsection{Portfolio Construction}\label{ssec:matching}
We construct the portfolio of pairs of stocks  using a given universe of stocks. 
Let the pair $a$ be given by stocks $i$ and $j$.  For each pair $a$ in the set of all stock pairs, we perform a linear regression of the log-prices  of the stocks using historic data:

\begin{align}
    \log(p_{jt}) & = \mu_a  + \beta_a \log(p_{it}) + \epsilon_{at}. \label{eq:regression} 
\end{align}
This regression allows us to estimate  the regression coefficients $\beta_a$ and $\mu_a$ for the pair, along with the variance of the residual $\epsilon_{at}$, which we denote as $\sigma^2_a$.  The residual of the regression is the pair spread that we use for trading.  To assess whether the pair is cointegrated, we  conduct a single-lag Augmented Dickey-Fuller (ADF) test on the spread, calculating the resultant t-statistic and p-value.  This process is repeated for each pair.

After computing the t-statistic of the ADF test for each pair, we construct a graph where the nodes are the stocks, the edges are the pairs, and the edge weights are the negative of the t-statistic.  We refer to this as the \textit{pairs graph}.  We use the negative of the t-statistic for the edge weight so that pairs that are more strongly cointegrated have a larger edge weight.  The pairs graph is fully connected and represents all the possible pairs and their level of cointegration.  We would like to select pairs that have a very negative t-statistic, indicating strong cointegration.  This corresponds to edges in the graph with large weight.  A portfolio consisting of a set of pairs can be represented as the subgraph of the pairs graph induced by the edges corresponding to these pairs.  We refer to this as the \textit{portfolio graph}.  Two-hop paths in the portfolio graph represent pairs of pairs that share a stock.  We would like to eliminate these types of pairs to reduce the portfolio variance, while also choosing pairs with large weight so they are likely to be cointegrated and give a positive return. This translates to a portfolio graph where there are no two-hop paths and the sum of the edge weights in maximized.  Such a graph can be obtained by finding a maximum weight matching on the pairs graph.

A matching is defined as a subgraph of a given graph in which each node is connected by at most one edge, ensuring that no two edges share a node. Among various types of matchings, a maximum weight matching is one where the total sum weights of the edges are maximized. There are a variety of algorithms that exist for finding maximum weight matchings on graphs.  Further details on such algorithms can be found in \cite{edmonds} and \cite{galil}.


To select the pairs for the matching portfolio, we construct a maximum weight matching on the pairs graph.  Note that if there are $n$ stocks, only $n/2$ pairs will be selected (assuming $n$ is even).  Therefore, there is an upper bound on the number of pairs in a matching portfolio.  We also note that while two stocks may form a highly cointegrated pair, this pair may not be selected because the maximum weight matching might select one or both of the stocks for two other pairs, so long as the sum of the weights is maximized.  This shows the potential complexity in constructing the matching portfolio.

\begin{figure}[h]
\centering
        \includegraphics[width=0.8\textwidth]{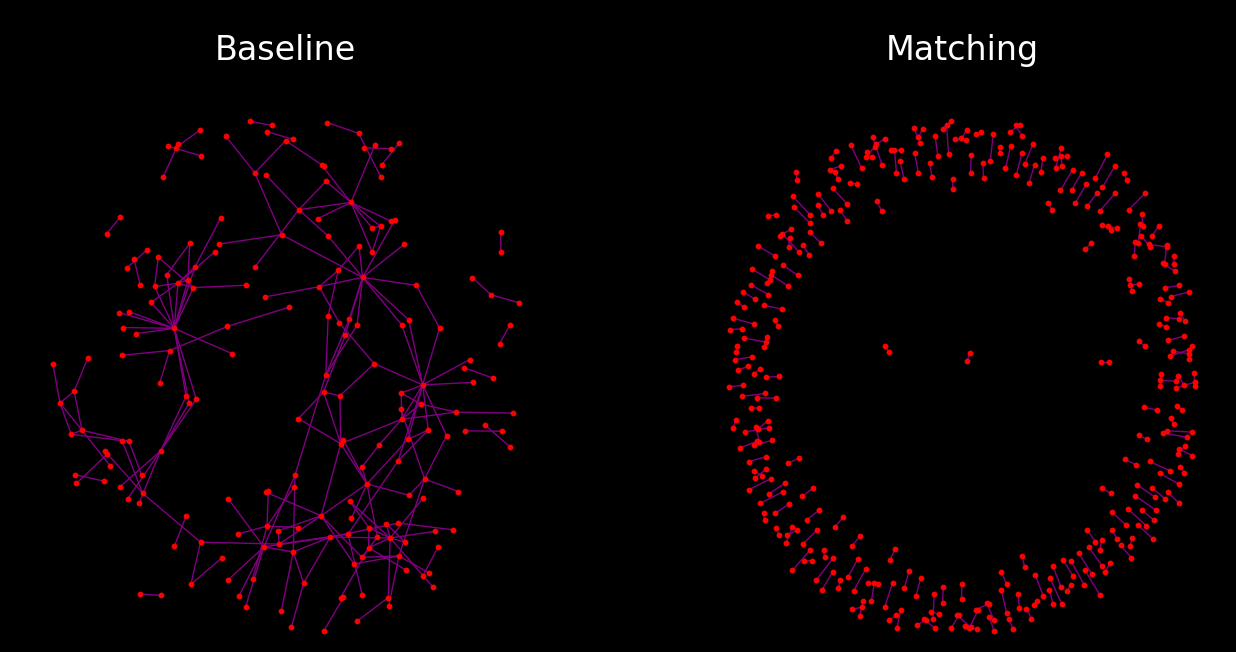}
        \caption{Portfolio graphs for (left) a baseline portfolio and (right) a matching portfolio, each with equal number of edges.  }
    \label{fig:graphs}
\end{figure}

One may be tempted to think that simply selecting the pairs with the most negative t-statistic would be a good way to build a portfolio.  This is the baseline portfolio discussed in Section \ref{ssec:sharpe}.  However, if this approach was taken, the resulting portfolio would have a large number of pairs of pairs sharing stocks.  To illustrate this, we consider two portfolios of equal size.  One is a baseline portfolio formed from the pairs with the most negative t-statistic (and subsequently lowest p-value) based on two years of data from the S\&P 500 from November 3rd, 2015 to November 3rd, 2017. 
The second is a matching portfolio constructed using the same pairs data.  Both portfolios have the same number of pairs, but have very different structures.   We show the two graphs in Figure \ref{fig:graphs}.  We can clearly see that the baseline portfolio has many two-hop paths, even though it has the same number of edges as the matching graph.  In fact,there is a stock in the baseline portfolio that is a part of 15 pairs.  In contrast, the matching portfolio contains only isolated edges and no two-hop paths.  All stocks in this portfolio are only in one pair.  This shows that by just selecting highly cointegrated pairs without utilizing any matching algorithm, one is likely to end up with a portfolio with a large variance due to the presence of many pairs sharing stocks.

\subsection{Pairs Trading}\label{ssec:trading}
After selecting pairs for the portfolio, we trade them throughout the duration of the trading period. The trading strategy for each pair in the portfolio is as follows. On each trading day, we calculate the regression of the log-prices as per equation \eqref{eq:regression}, using a fixed-size historical data window leading up to that day. We then obtain the day's residual from this regression and standardize it by dividing by the regression's estimated variance, obtaining a z-scored spread. To limit the influence of extreme outliers, this z-scored spread is winsorized at the -3 and 3 thresholds. The trading signal, as defined by equation \eqref{eq:signal}, employs this winsorized spread in lieu of the original spread. We select a specific value for $k$ to generate the signal. The trading methodology based on this signal adheres to the framework outlined in Section \ref{sec:theory} with a small modification.  The regression produces a coefficient $\beta_a$ which relates returns of stocks $i$ and $j$.  Typically this value is near one, but if it deviates from one, we must modify how we allocate capital to the two stocks.   For a pair $a$ comprising stocks $i$ and $j$, where the residual $\epsilon_{at}$ is calculated as per equation \eqref{eq:regression}, a signal of +1 prompts the purchase of stock $i$ with $\beta_a$ dollars  and the short-selling of one dollar of stock $j$. Conversely, a signal of -1 leads to the purchase of one dollar of stock  of $j$ and the short-selling of $\beta_a$ dollars of stock $i$. A zero signal indicates no position is taken in either stock.

One potential draw-back of this z-scored signal is that the underlying regression may be very sensitive to outliers.  For example, with a very large estimated standard deviation (due to a single outlier) in the denominator, the z-score may not qualify for an otherwise appropriate pair trade. To overcome this sensitivity, 
we propose a \textit{q-score} that relies on only quantiles of the residual distribution. Let $\tau_{k} (\epsilon_{at} )$ be the $k^{th}$ quantile of the residuals of the regression at time $t$. Then we define the q-score of the residual for pair $a$ at time $t$ as

\begin{equation}
\label{eq:qscore}
q_{at} = \frac{\epsilon_{at} - \tau_{50} (\epsilon_{at} )}{\tau_{75} (\epsilon_{at} ) - \tau_{25} (\epsilon_{at} )}
\end{equation}
The numerator subtracts the median from the spread at time $t$, and the denominator is the difference between the 75th and 25th quantiles of the residuals. Because only quantiles are used, the q-score provides a measure of deviation that is robust to outliers.  

Another modification we make to the signal is how we determine the weight of the position in the portfolio.  The unnormalized position weight is the absolute value of the signal.  For the z-score signal given by equation \eqref{eq:signal}, each pair had an equal weight.  However, if the spread is very far from zero, this presents an opportunity for a higher return when the spread mean-reverts.  Therefore, we would want to have a larger weight in the portfolio for pairs whose residual is far from zero.  To achieve this, we define the following modified signal based on the q-score:

\begin{align}
    S^q_{at} =   \paranth{-\mathbbm{1}\curly{q_{at}\leq 0}   
                + \mathbbm{1}\curly{q_{at}\geq 0}}[|q_{at}|]
    \label{eq:signal_q}.
\end{align}
where $[|q_{at}|]$ means round the absolute value of $q_{at}$ to the nearest integer.  The rounding prevents any trade from occurring if the q-score is too small (in particular if it is between -0.5 and 0.5).  This was achieved in the z-score by explicit selection of the threshold $k$.  In addition to the prevention of trading on small spreads, we also find that the rounding leads to good performance in the trading simulations shown in Section \ref{sec:results}.  This q-score based signal has the same long and short behavior as the z-score signal, except now the absolute value of the signal is proportional to the absolute value of the q-score.  This allows for greater weight on pairs where the spread is far from zero.

%

\section{Results}\label{sec:results}
In this section we discuss the implementation of our matching-based pairs trading strategy and the associated performance results. First, we define the various trading strategies and performance metrics we utilize. Then, we describe the details of our trading simulations.  Finally, we present the performance results. The relevant code and data utilized in this section can be found in \citep{github}.

We use a combination of metrics measured before (ex-ante) and after (ex-post) realizing the trading performance.  Ex-post metrics concern the actual performance of the portfolio, while  ex-ante metrics help us understand the differences in the properties of the portfolios for different trading strategies.  For ex-post metrics, we consider the cumulative returns of the strategies benchmarked to the S\&P 500.  We also consider risk-adjusted measures of the returns such as the Sharpe ratio and Sortino ratio. The Sharpe ratio measures the ratio between mean returns and standard deviation of returns \citep{sharpe}.  The Sortino ratio measures the ratio between mean returns and standard deviation of negative returns, and captures downside risk \citep{sortino}. For both measures, a higher value indicates a higher return per unit of risk.  Another important measure we consider is the drawdown, which is the largest drop in cumulative returns between peak and trough over the investigated time period. This metric is commonly used by fund managers to account for periodic under-performance  and recovery duration \citep{choi}. We also consider the single worst and best day of returns.  

For ex-ante metrics, we focus on asset turnover, retention, and concentration within the portfolio. Asset turnover is critical because frequent entries and exits from positions may incur substantial trading costs. Retention addresses the portfolio's diversity in terms of its holdings over time, aiming for low retention to reduce the risk associated with holding a particular pair for an extended period. Concentration measures the maximum frequency of a single stock's appearance in the portfolio, where a high concentration level could significantly increase the portfolio's vulnerability to single stock risk.

\subsection{Trading Simulation}\label{ssec:sim}
 Our trading simulations utilize historical price data for stocks on the S\&P 500 \citep{spglobal}  between January 2017 and May 2023 adjusted for both dividends and stock splits \citep{yahoo}. We update the pairs in the portfolio once a month, using whatever pair selection method we are testing.  Then for the duration of the month we trade those pairs using one of the approaches described in Section \ref{ssec:trading}.  To compute the pair regressions for both selection and trading we use two years of historical data.  This window size is commonly used in the pairs trading literature \citep{gatev, vidyamurthy}. Daily trading is done using a sliding two-year window to compute the spreads.  

To account for trading costs, which can be significant when considering market impact and transaction costs \citep{dotc}, we use the following approach.
We model the annual transaction costs for each stock at 1\%, translating to a daily fee of $1\%/252\text{ days} = 0.00397\%$, consistent with the findings in \cite{secfees}.  When trading occurs for a pair, the daily net returns are the gross returns minus this transaction fee.

We utilize two different approaches to select the pairs for the portfolio.  One is the matching approach described in Section \ref{ssec:matching} which utilizes a maximum weight matching.  The second is a baseline approach where we sort the pairs by their p-values, and select the top 250 pairs for the feasible pair universe. By selecting only 250 pairs, we are able to compare the matching and baseline approaches while controlling for the number of pairs in the portfolio. For trading the pairs, we compare the z-score signal in  equation \ref{eq:signal} and the q-score signal in equation \ref{eq:signal_q}.

\subsection{Returns}
\begin{figure}
    \centering
    \begin{subfigure}{0.49\textwidth}
        \centering
        \includegraphics[width=\linewidth]{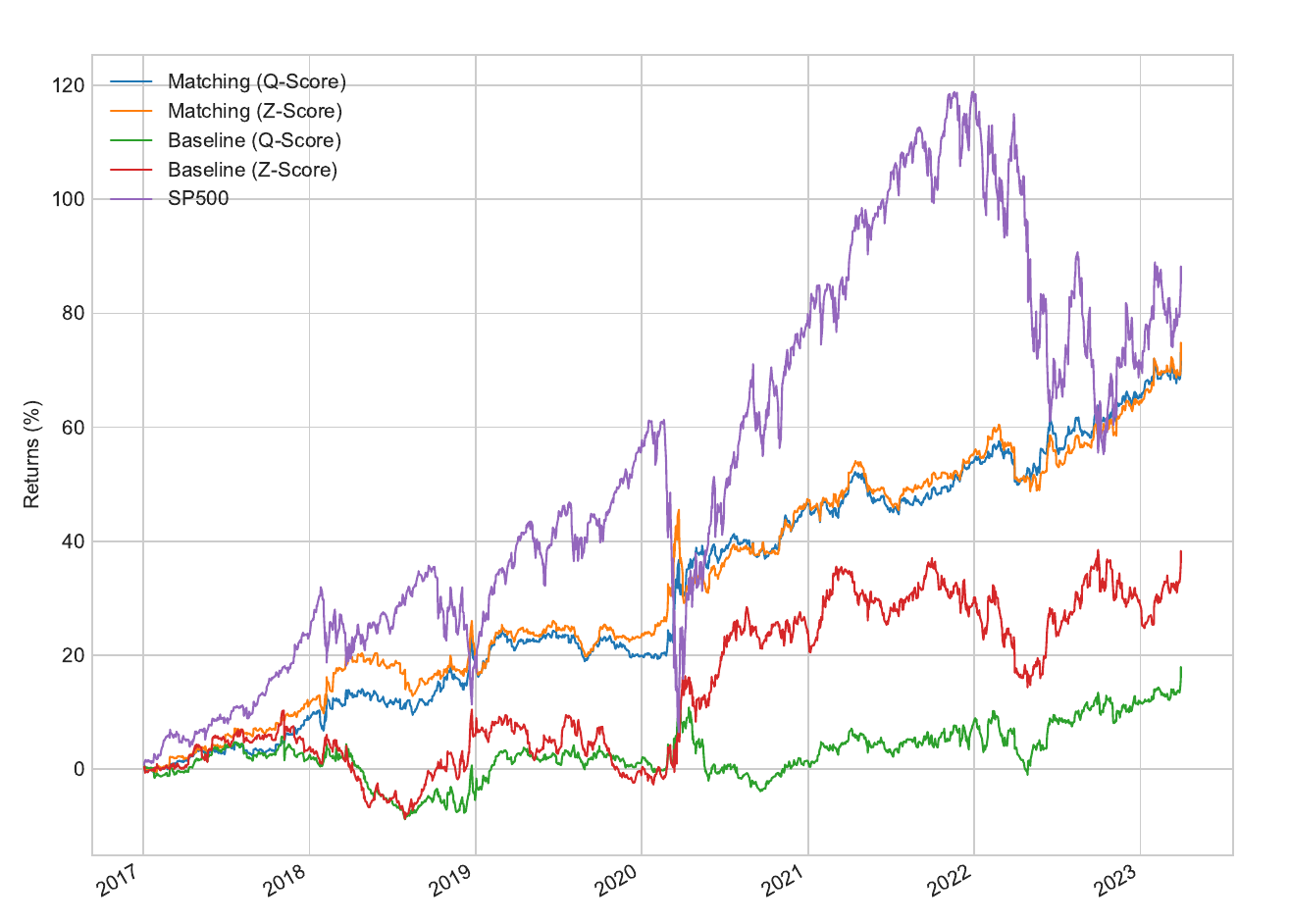}
    \end{subfigure}
    \hfill
    \begin{subfigure}{0.49\textwidth}
        \centering
        \includegraphics[width=\linewidth]{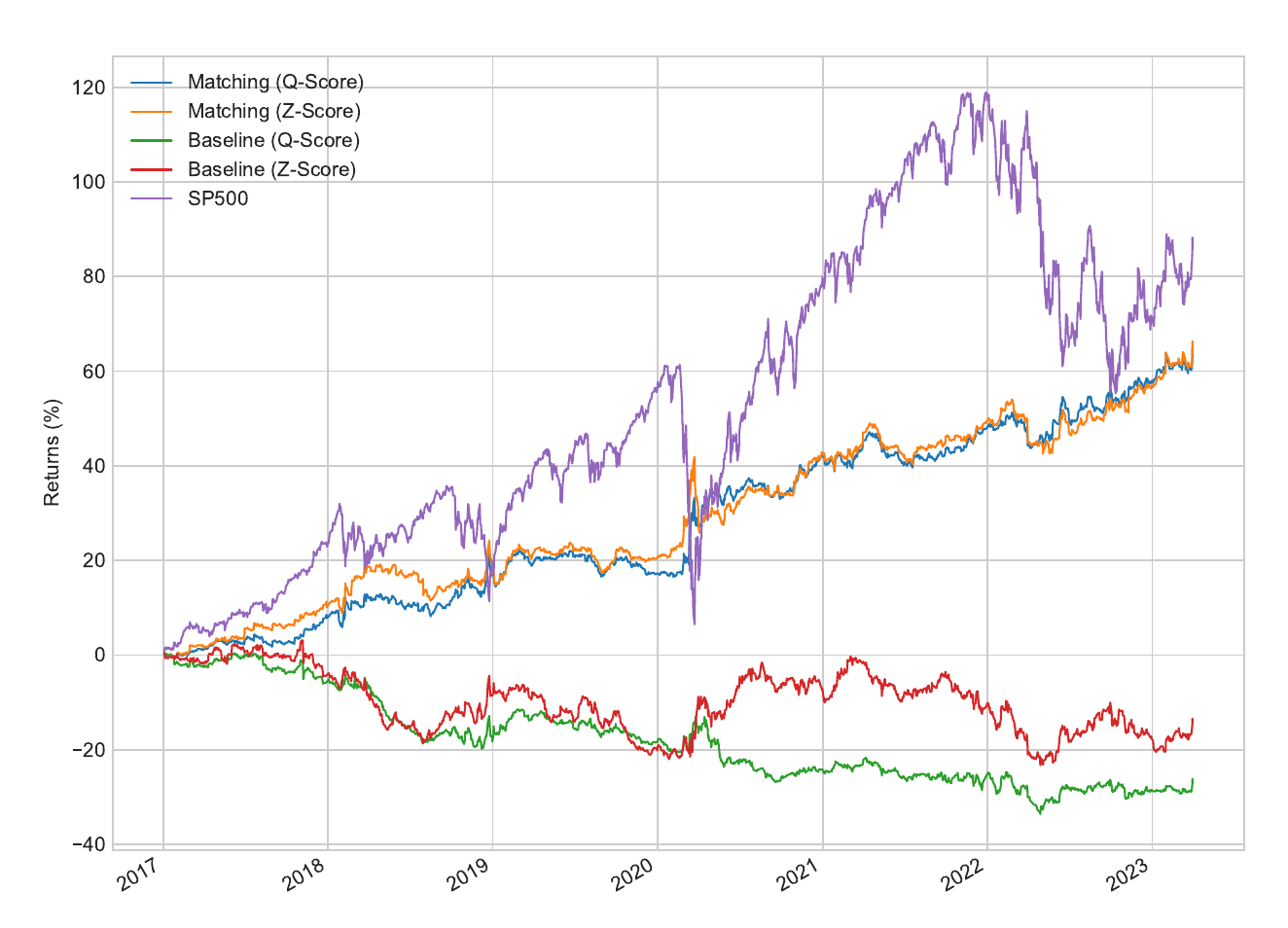}
    \end{subfigure}
    \caption{(right) Gross  and (left)  net cumulative returns (accounting for fixed transaction costs) of the different trading strategies.}
    \label{fig:cum} 
\end{figure}

We begin by analyzing the returns of the different pairs trading strategies.  We denote the pairs selection strategies as either M (matching) or B (baseline), and we denote the trading signal as Z (z-score) or Q (q-score).  We plot the cumulative gross and net returns in Figure \ref{fig:cum}.  We see that all of the strategies had positive gross returns. The baseline approaches (BQ and BZ) returned 18\% and 38\%, while the matching approaches (MQ and MZ) returned 75\% and 74\%.  For reference, in this period the S\&P 500 returned 88\%.  We also plot the net returns (calculated using the procedure described in Section \ref{ssec:sim}) in Figure \ref{fig:cum}.  The matching approaches still outperform the baseline approaches, which end up having negative returns when factoring in transaction costs.  Overall, the matching approaches returned near 66\%, while the baseline approaches returns -26.15\% and  -13.40\%.  This exemplifies that turnover considerations are essential to the design of a pairs trading strategy since the baseline actually becomes unprofitable when realistic transaction costs are taken into account.

We summarize a variety of performance metrics for the different strategies in  Tables \ref{table:performance_metrics_gross} and \ref{table:performance_metrics_net}.  The net annualized returns of the matching-based strategies are  7.64\% and 7.25\%, slightly below the S\&P 500's value of 10.79\%.  However, while the matching approaches achieved slightly lower returns compared to the market, they stood out due to their superior risk-adjusted performance.  We see that the S\&P 500 had a Sharpe ratio of 0.59, while matching approaches both had a Sharpe ratio of 1.23 and Sortino ratios of 1.69 and 1.78 based on gross returns. Furthermore, the S\&P 500 experienced much larger single-day losses. For example, the S\&P 500 experienced a -11\% return on a single day, where as the matching approaches never exceeded -5\%. On a cumulative basis, the S\&P 500 experienced a -19.49\%  drawdown, while all pairs trading approaches had drawdowns that did not exceed -8.48\%.  This analysis indicates that matching-based pairs trading not only offers returns comparable to the market, but does so with considerably lower risk, even surpassing market performance in bearish years like 2022.  Interestingly, the Sharpe ratios for both the baseline and matching portfolios closely align with the theoretical values presented in Table \ref{table:params}, underscoring the accuracy of our theoretical framework.

We also see differences in the performance of the two trading signals.  While the q-score signal has a slightly lower annualized return than the z-score signal,  it does mitigate the minimum single day return and drawdown.  This indicates that the q-score signal may offer protection against large losses at the cost of slightly lower returns.


\begin{table}
    \centering
    \caption{Performance metrics for the different trading strategies based on gross returns.}
    \label{table:performance_metrics_gross}
    \begin{tabular}{lrrrrr}
        \toprule
        {} &     MQ &     MZ &     BQ &     BZ &  S\&P 500 \\
        \midrule
        Gross Sharpe ratio            &   1.23 &   1.23 &  0.34 &  0.48 &   0.59 \\
        Gross Sortino ratio           &   1.78 &   1.69 &  0.44 &  0.67 &   0.70 \\
        Gross cumulative returns (\%) &  73.65 &  74.87 & 17.96 & 38.30 &  88.25 \\
        Gross annualized returns (\%)   &  8.40 &  8.42 & 2.49 &  5.13 &  10.79 \\
        Minimum gross single day return (\%)           &  -3.81 &  -4.16 &  -3.93 &  -4.63 &  -10.94 \\
        Maximum gross single day return (\%)         &   4.20 &   4.13 &   3.49 &   4.99 &   9.06 \\
        Skew              &   1.12 &   0.83 &  -0.28 &   0.04 &  -0.55 \\
        Drawdown (\%)          &  -7.99 &  -8.30 &  -6.94 &  -8.48 & -19.49 \\
        \bottomrule
    \end{tabular}
\end{table}

\begin{table}
    \centering
    \caption{Performance metrics for the different trading strategies based on net returns.}
    \label{table:performance_metrics_net}
    \begin{tabular}{lrrrrr}
        \toprule
        {} &     MQ &     MZ &     BQ &     BZ &  S\&P 500 \\
        \midrule
        Net Sharpe ratio            &   1.12 &   1.12 &  -0.49 &  -0.13 &   0.59 \\
        Net Sortino ratio           &   1.63 &   1.55 &  -0.64 &  -0.18 &   0.70 \\
        Net cumulative returns (\%) &  64.73 &  66.34 & -26.15 & -13.40 &  88.25 \\
        Net annualized returns (\%)   &  7.64 &  7.25 & -4.11 &  -1.65 &  10.79 \\
        Minimum net single day return (\%)           &  -3.82 &  -4.17 &  -4.01 &  -4.67 &  -10.94 \\
        Maximum net single day return (\%)         &   4.18 &   4.13 &   4.90 &   3.44 &   9.06 \\
        Skew              &   1.12 &   0.82 &  -0.31 &   0.03 &  -0.55 \\
        Drawdown (\%)          &  -8.00 &  -8.30 &  -6.90 &  -8.40 & -19.49 \\
        \bottomrule
    \end{tabular}
\end{table}

We next considered the correlation of the gross returns  between the various trading strategies in Table \ref{table:corr}.  We find that the strategies show strong correlations with each other, but all have a slightly negative correlation with the S\&P 500.  This may indicate that pairs trading strategies behave contrarian to market growth, and can actually be profitable when the market is falling. The S\&P 500 had lower risk adjusted returns than the matching-based strategies. Furthermore, during periods of drawdown in the S\&P 500 such as COVID-19 and the Russian invasion of Ukraine, amongst others, the matching-based strategies outperformed the market. For example, during March 2020 the S\&P 500 returned -35\% while the matching approach returned 15\%. During another bearish period in 2022, the S\&P 500 returned -18\% while the matching approach returned 10\%.

\begin{table}[h]
    \centering
    \begin{tabular}{lrrrrr}
        \toprule
        {} &    MQ &    MZ &    BQ &    BZ &  S\&P 500 \\
        \midrule
        MQ    &  1.00 &  0.91 &  0.50 &  0.35 &  -0.27 \\
        MZ    &  0.91 &  1.00 &  0.51 &  0.37 &  -0.25 \\
        BQ    &  0.50 &  0.51 &  1.00 &  0.68 &  -0.17 \\
        BZ    &  0.35 &  0.37 &  0.68 &  1.00 &  -0.14 \\
        S\&P 500 & -0.27 & -0.25 & -0.17 & -0.14 &   1.00 \\
        \bottomrule
    \end{tabular}
    \caption{The correlation of returns between four different trading strategies and the S\&P 500.}
    \label{table:corr}
\end{table}





\subsection{Asset Turnover and Retention}
\begin{figure}[ht]
\centering
        \includegraphics[width=0.7\textwidth]{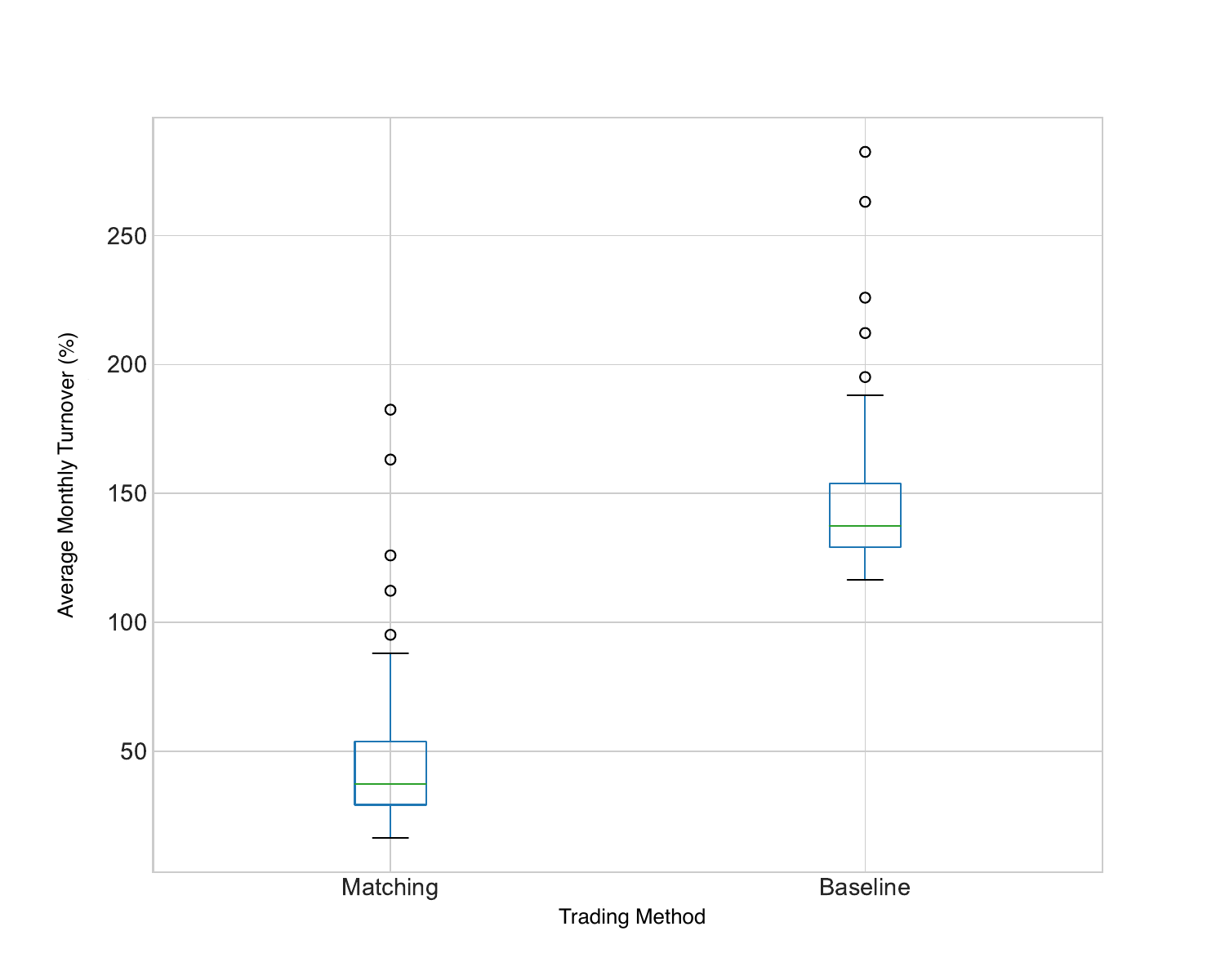}
        \caption{Boxplot of the average monthly turnover for the matching and baseline portfolios using the q-score signal. }
    \label{turnover}
\end{figure}

One important consideration in  executing trading strategies is the cost of entry and exit into a position. With excessive trading, execution costs increase, resulting in performance degradation. Consequently, investment managers prefer strategies with lower turnover that require fewer trades. We define \textit{turnover}, $\Delta(t)$ in time period $t$ as the sum of the absolute difference of positions  for each asset  in time periods $t-1$ and $t$ for each asset.  If we define $x_j(t)$ as the position at time $t$ in asset $j$ in portfolio $U$, the the turnover is 

\begin{equation}
\Delta(t) = \sum_{j \in U} |x_j (t) - x_j (t-1)|.
\end{equation}
We show the distribution of the average monthly turnover for the baseline and matching methods in Figure \ref{turnover} based on the q-score signal.  The matching method has a median turnover one third as large as the baseline method.   This higher turnover for the baseline method is a major factor in the reduction of net returns that we observed in Figure \ref{fig:cum}.  In contrast, the matching method trades much less frequently, is not penalized as severely by transaction costs, and ends up being profitable.

Another important consideration is the diversity of the portfolios over time.  In our trading strategies, we update the traded pairs on a monthly basis. Suppose for example, that the portfolio contained the same pairs each month. This may imply that the portfolio construction method is flawed because no real selection is being carried out. This also exposes the portfolio to greater risk from holding the pair too long. To quantify this concept of diversity, we measure the similarity of two portfolios using the Jaccard index \citep{jaccard}. Formally, for a given pair selection strategy $U$, we define the set of pairs in the portfolio at time period $t$ as $U(t)$.  We then define the  \textit{retention} at time $t$ as the Jaccard index between portfolios in consecutive update periods, which is given by
\begin{equation}
    J(U(t), U(t-1)) = \frac{|U(t) \cap U(t-1)|}{|U(t) \cup U(t-1)|}.
\end{equation}
This coefficient is equal to one if the two portfolios are identical and zero if they are completely disjoint. 
 
\begin{figure}[ht]
\centering
        \includegraphics[width=0.7\textwidth]{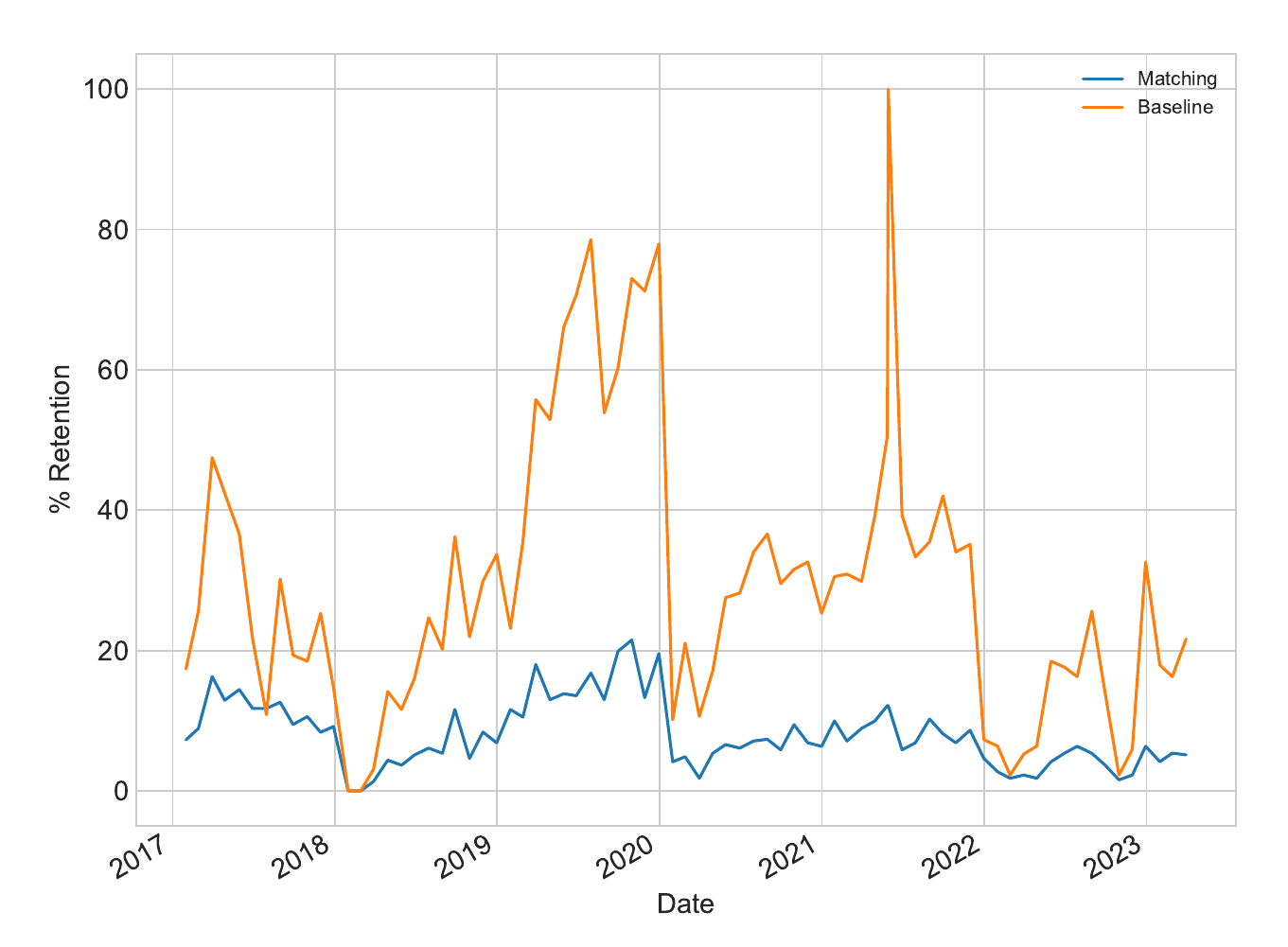}
        \caption{Plot of the monthly pair retention versus time for the baseline and matching portfolios.}
    \label{fig:retention}
\end{figure}

 We show the historical retention of the matching and baseline strategies in Figure \ref{fig:retention}.  We see that the matching method consistently has a low retention, implying a higher diversity than the baseline method. We also see that the retention fluctuates a great deal for the baseline method. This indicates that it is very inconsistent, with some periods having very stable portfolios while other periods having very different portfolios.

\subsection{Concentration}
\begin{figure}[ht]
\centering
        \includegraphics[width=0.7\textwidth]{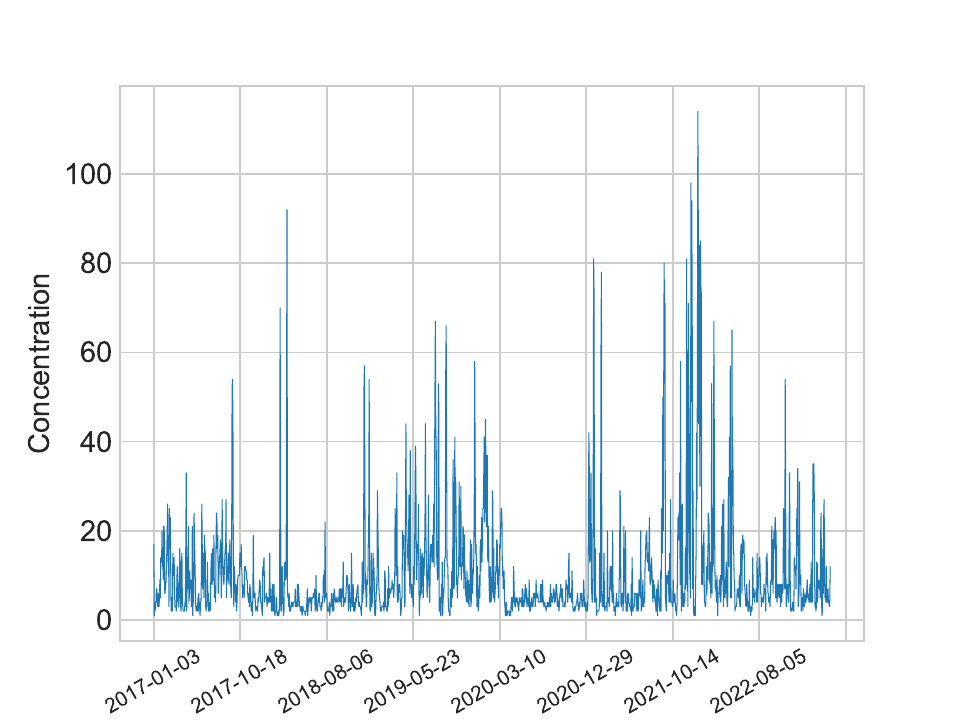}
        \caption{Plot of the single-stock concentration versus time for the baseline portfolio. }
    \label{fig:concentration}
\end{figure}

An additional benefit of the matching approach is that it prevents single-stock concentration, which is a concern for managing idiosyncratic risk \citep{goyal}. Single-stock risk is an important consideration within active portfolio management, and managers are often required the abide by  strict single-stock exposure constraints \citep{grinold}. In the baseline approach, it is possible that a specific stock is selected for multiple pairs, resulting in a significant increase in exposure to idiosyncratic risk. This exposure could be particularly high when the stock goes through a corporate action event such as earnings or management change, and every pair containing it would have an abnormal residual. With the matching approach on the other hand, each stock is selected only once, and the single-stock risk is limited to a single pair. To formalize this, we define the \textit{concentration} of a portfolio as the maximum number of pairs in which any single stock appears.  Within the context of the portfolio graph, this is the same as the maximum degree among all nodes in the graph.  The matching method has a concentration of one by construction.  We plot the concentration of the baseline method  in Figure \ref{fig:concentration}. The average concentration is 10.2, but the concentration fluctuates quite a bit and can exceed 100.  We see that the baseline method can often choose many pairs that share the same stock.  This is a result of the pairs selection criteria which is based on the p-value of a cointegration test.  The high concentration likely arises due to a small set of stocks having strong cointegration at any given time.

\section{Conclusion}
In this study, we have demonstrated the advantages of employing a matching-based selection method for pairs trading, in contrast to traditional approaches that rely solely on the significance of cointegration tests for pair selection. The latter can result in portfolios overly concentrated in a limited number of stocks, which elevates portfolio variance and diminishes risk-adjusted returns.
Our analysis includes analytical expressions for the mean and variance of returns for portfolios composed of both cointegrated and non-cointegrated pairs. We leveraged these expressions to illustrate that a matching-based approach significantly mitigates portfolio variance when compared to conventional strategies. Trading simulations corroborate our theoretical findings, revealing that matching-based pairs trading not only delivers robust risk-adjusted performance but also achieves annualized returns comparable to the S\&P 500 with a superior Sharpe ratio.

A unique aspect of our methodology is the application of graph theory to the selection of stock pairs. By conceptualizing pairs portfolios as graphs, we unveil strategies to decrease variance through optimal matchings. Our method assigns the t-statistic from a cointegration statistical test as the weight for the edges. Furthermore, while we opted for a maximum weight matching, exploring alternative weighting schemes or selection criteria, such as excluding pairs with weak cointegration, presents intriguing avenues for enhancing the selection process.  One could also explore a variant of pairs trading that encompasses cointegration among multiple stocks, transforming the concept of a pair into a tuple and evolving the portfolio graph into a hypergraph. Our matching-based strategy can very naturally to extend to this hypergraph version of pairs trading, facilitating the creation of low variance portfolios for more complex notions of pairs. Therefore, this graph-theoretical approach to pairs trading paves the way for numerous innovative modifications to the pairs trading methodology.

\bibliography{bibs}

\begin{thebibliography}{26}
\providecommand{\natexlab}[1]{#1}
\providecommand{\url}[1]{\texttt{#1}}
\expandafter\ifx\csname urlstyle\endcsname\relax
  \providecommand{\doi}[1]{doi: #1}\else
  \providecommand{\doi}{doi: \begingroup \urlstyle{rm}\Url}\fi

\bibitem[Avellaneda~M.(2010)]{avellaneda}
Lee~J.H. Avellaneda~M.
\newblock Statistical arbitrage in the us equities market.
\newblock \emph{Journal of Derivatives and Hedge Funds}, 2010.

\bibitem[Bollobás(2001)]{bollobas2001random}
Béla Bollobás.
\newblock \emph{Random Graphs}.
\newblock Cambridge University Press, Cambridge, 2 edition, 2001.

\bibitem[Botha(2013)]{botha}
et.~al. Botha.
\newblock Trading strategies with copulas.
\newblock \emph{Quantitative Finance}, 2013.

\bibitem[Choi(2021)]{choi}
Jaehyung Choi.
\newblock Maximum drawdown, recovery, and momentum.
\newblock \emph{Journal of Risk and Financial Management}, 14\penalty0 (11):\penalty0 542, 2021.

\bibitem[Cummins~M.(2012)]{cummins}
Bucca~A. Cummins~M.
\newblock Quantitative spread trading on crude oil and refined products markets.
\newblock \emph{Quantitative Finance}, 2012.

\bibitem[Do and Faff(2012)]{dotc}
Binh Do and Robert Faff.
\newblock Are pairs trading profits robust to trading costs?
\newblock \emph{Journal of Financial Research}, 35\penalty0 (2):\penalty0 261--287, 2012.

\bibitem[Edmonds(1965)]{edmonds}
Jack Edmonds.
\newblock Paths, trees, and flowers.
\newblock \emph{Canadian Journal of mathematics}, 17:\penalty0 449--467, 1965.

\bibitem[Elliott~R.J.(2004)]{elliott}
et~al. Elliott~R.J.
\newblock Pairs trading.
\newblock \emph{Quantitative Finance}, 5:\penalty0 271--276, 2004.

\bibitem[Enders(2004)]{walter}
W.~Enders.
\newblock Cointegration and error-correction models.
\newblock \emph{Wiley}, 2004.

\bibitem[Galil(1986)]{galil}
Z.~Galil.
\newblock Efficient algorithms for finding maximum matching in graphs.
\newblock \emph{ACM Computing Surveys}, 1986.

\bibitem[Gatev~E.(2006)]{gatev}
Rouwenhorst~K.G. Gatev~E., Goetzmann~W.N.
\newblock Pairs trading: Performance of a relative value arbitrage rule.
\newblock \emph{The Review of Financial Studies}, 2006.

\bibitem[GitHub.(2023)]{github}
Qureshi GitHub.
\newblock Github.
\newblock \url{https://github.com/kai-trading-bot/pair/}, 2023.
\newblock Accessed: 2023-04-30.

\bibitem[Goyal~A.(2003)]{goyal}
Santa Clara~P. Goyal~A.
\newblock Idiosyncratic risk matters!
\newblock \emph{Journal of Finance}, 2003.

\bibitem[Grinold~R.(2000)]{grinold}
Kahn~R.N. Grinold~R.
\newblock \emph{Active Portfolio Management}.
\newblock McGraw-Hill, 2000.

\bibitem[Huck(2009)]{huck}
Huck.
\newblock Pairs selection and outranking: An application to the s and p 100 index.
\newblock \emph{European Journal of Operational Research}, 2009.

\bibitem[Huck~N.(2015)]{huck2}
Afawubo~K. Huck~N.
\newblock Pairs trading and selection methods: is cointegration superior?
\newblock \emph{Applied Economics}, 2015.

\bibitem[Jaccard(1901)]{jaccard}
P.~Jaccard.
\newblock Étude comparative de la distribution florale dans une portion des alpes et des jura.
\newblock \emph{Bulletin de la Société vaudoise des sciences naturelles}, 37, 1901.

\bibitem[Jurek~J.(2007)]{jurek}
Yang~H. Jurek~J.
\newblock Dynamic portfolio selection in arbitrage.
\newblock \emph{EFA 2006 Meetings Paper}, 2007.

\bibitem[Liew(2013)]{liew}
Liew.
\newblock Pairs trading: A copula approach.
\newblock \emph{Journal of Derivatives and Hedge Funds}, 2013.

\bibitem[Liu~J.(2012)]{liu}
Timmermann~A. Liu~J.
\newblock Optimal convergence trade strategies.
\newblock \emph{Review of Financial Studies}, 2012.

\bibitem[SEC(2019)]{secfees}
SEC.
\newblock How fees and expenses affect your investment portfolio, 2019.
\newblock URL \url{https://www.sec.gov/oiea/investor-alerts-bulletins/ib-fees-expenses}.

\bibitem[Sharpe(1994)]{sharpe}
William~F Sharpe.
\newblock The sharpe ratio, the journal of portfolio management.
\newblock \emph{Stanfold University, Fall}, 1994.

\bibitem[Sortino and Price(1994)]{sortino}
Frank~A Sortino and Lee~N Price.
\newblock Performance measurement in a downside risk framework.
\newblock \emph{the Journal of Investing}, 3\penalty0 (3):\penalty0 59--64, 1994.

\bibitem[SPGlobal(2020)]{spglobal}
SPGlobal.
\newblock Spglobal.
\newblock \url{https://www.spglobal.com/en/}, 2020.
\newblock Accessed: 2023-04-30.

\bibitem[Vidyamurthy(2004)]{vidyamurthy}
G.~Vidyamurthy.
\newblock \emph{Pairs Trading, Quantitative Methods and Analysis}.
\newblock Wiley, 2004.

\bibitem[Yahoo(2023)]{yahoo}
Yahoo.
\newblock Yahoo finance.
\newblock \url{https://finance.yahoo.com/}, 2023.
\newblock Accessed: 2023-04-30.

\end{thebibliography}
\bibliographystyle{plainnat}

\section{Appendix}
\subsection{Proof of Theorem \ref{thm:cointegrated_mean_var}}
We prove here the results of Theorem \ref{thm:cointegrated_mean_var} for the mean and variance of a cointegrated pair.
We assume the cointegrated pair consists of stocks 1 and 2, and using the definition of the spread and log-normal price model for stocks 1, we can write the return of each stocks at time $t+1$ as
\begin{align*}
    r_{1,t+1} & = e^{\mu_1+\delta_{1,t+1}}-1\\
    r_{2,t+1} & = e^{\mu_1+\delta_{1,t+1}+\epsilon_{t+1}-\epsilon_{t}}-1.
\end{align*}
The trading signal is given  by equation \eqref{eq:signal}, which we reproduce here for convenience:
\begin{align*}
            S_t =  \mathbbm{1}\curly{\epsilon_t \leq -k\sigma} -   \mathbbm{1}\curly{\epsilon_t \geq k\sigma}.
\end{align*}
The return of the spread is $r_{t+1} = S_t(r_{2,t+1} - r_{1,t+1})$, which  we can write as 
\begin{align}
    r_{t+1} = S_te^{\mu_1+\delta_{1,t+1}}\paranth{e^{\epsilon_{t+1} -\epsilon_t} -1}.\label{eq:returns_coint}
\end{align}
We will now calculate the mean and variance of this quantity. We begin with the mean.  We can easily take the expectation with respect to $\delta_{1,t+1}$ and $\epsilon_{t+1}$ to obtain
$$ \mathbf E[r_{t+1}|\epsilon_t]= S_t e^{\mu_1 + \sigma_1^2/2}(e^{\sigma^2/2-\epsilon_t}-1).  $$
The expectation over $\epsilon_t$ is more complicated as both the signal and the term in the parentheses depend upon it.  We start by observing that $\mathbf E[S_t]=0$.  Applying this we obtain
$$ \mathbf E[r_{t+1}]= e^{\mu_1 + \sigma_1^2/2+\sigma^2/2}\mathbf E[S_te^{-\epsilon_t}].$$
If we make the substitution $\epsilon_t = \sigma z$, we can write the expectation as
$$ \mathbf E[S_te^{-\epsilon_t}] = \frac{1}{\sqrt{2\pi}}\int^{-k} 
e^{-\sigma z-z^2/2}dz
-\frac{1}{\sqrt{2\pi}}\int_{k} 
e^{-\sigma z-z^2/2}dz.$$
To simplify this expression, we complete the square in the exponent to obtain
$$ \mathbf E[S_te^{-\epsilon_t}] = \frac{e^{\sigma^2/2}}{\sqrt{2\pi}}\paranth{\int^{-k} 
e^{-(z+\sigma)^2/2}dz-
\int_{k} 
e^{-(z+\sigma)^2/2}dz}$$
which can be more easily expressed as
\begin{align}
\mathbf E[S_te^{-\epsilon_t}] = e^{\sigma^2/2}\paranth{\mathbf P(Z\leq -k+\sigma) - \mathbf P(Z\geq k+\sigma)}\label{eq:sexp}.
\end{align}
where $Z$ is a standard normal random variable with mean $0$ and unit variance.  Substituting this into the expression for the mean returns we obtain 
\begin{align*}
   \mathbf E[r_{t+1}] = e^{\mu_1+\sigma_1^2/2+\sigma^2}\paranth{\mathbf P(Z\leq -k+\sigma) - \mathbf P(Z\geq k+\sigma)}. 
\end{align*}
We can further simplify this by using the symmetry of $Z$ to obtain
\begin{align*}
   \mathbf E[r_{t+1}] = e^{\mu_1+\sigma_1^2/2+\sigma^2}
                        \mathbf P(k-\sigma\leq Z\leq k+\sigma) . 
\end{align*}
To compute the variance, we begin by analyzing the square pair returns.  We can write this as
\begin{align*}
(r_{t+1})^2 &= e^{2\mu_1 + 2\delta_{1,t+1}}S_t^2\paranth{e^{2\epsilon_{t+1}-2\epsilon_t}
                                                -2e^{\epsilon_{t+1}-\epsilon_t}+1}
\end{align*}
If we take the expectation over $\delta_{1,t+1}$ and $\epsilon_{t+1}$ we obtain
\begin{align*}
\mathbf E[(r_{t+1})^2] &= e^{2\mu_1+2\sigma_1^2}\paranth{              
                    e^{2\sigma^2}\mathbf E[S_t^2e^{-2\epsilon_t}]
                    -2e^{\sigma^2/2}\mathbf E[S_t^2e^{-\epsilon_t}]
                    +\mathbf E[S_t^2]}.
\end{align*}
We now simplify the different expectations in this expression.  First, we consider the simplest term involving $S_t^2$.  We can express this as
\begin{align*}
    \mathbf E[S_t^2] & = \mathbf E[\mathbbm{1}\curly{\epsilon_t \leq -k\sigma}] + \mathbf E[\mathbbm{1}\curly{\epsilon_t \geq k\sigma}] 
    -2 \mathbf E[\mathbbm{1}\curly{\epsilon_t \leq -k\sigma}] \mathbf E[\mathbbm{1}\curly{\epsilon_t \geq k\sigma}]\\
    & = \mathbf E[\mathbbm{1}\curly{\epsilon_t \leq -k\sigma}] + \mathbf E[\mathbbm{1}\curly{\epsilon_t \geq k\sigma}] \\
    & = \mathbf P(Z\leq -k) +\mathbf P(Z\geq k)\\
    & = 2\mathbf P(Z\geq k)
\end{align*}
where $Z$ is a standard normal random variable.

The other expectations have a similar structure, so we will derive a simplified expression for a more general quantity. We define $\epsilon_t=\sigma z$ and choose a constant $c>0$.  Then  we have
\begin{align*}
    \mathbf E[e^{-c\epsilon_t}S_t^2] & = \frac{1}{\sigma\sqrt{2\pi}} \paranth{
    \int^{-k\sigma}e^{-c\epsilon_t-\epsilon_t^2/(2\sigma^2)}d\epsilon_t
    +\int_{k\sigma}e^{-c\epsilon_t-\epsilon_t^2/(2\sigma^2)}d\epsilon_t
    }\\
    & = \frac{1}{\sqrt{2\pi}} \paranth{
    \int^{-k}e^{-c\sigma z-z^2/2}dz
    +\int_{k}e^{-c\sigma z-z^2/2}dz
    }.
\end{align*}
As we did when calculating the mean returns, we complete the square in the exponent to obtain
$$ \mathbf E[e^{-c\epsilon_t}S_t^2] = \frac{e^{c^2\sigma^2/2}}{\sqrt{2\pi}}\paranth{\int^{-k} 
e^{-(z+c\sigma)^2/2}dz
+\int_{k} e^{-(z+c\sigma)^2/2}dz}.$$
This can be simplified to
$$ \mathbf E[S_t^2e^{-c\epsilon_t}] = e^{c^2\sigma^2/2}\paranth{\mathbf P(Z\leq -k+c\sigma) + \mathbf P(Z\geq k+c\sigma)}
$$
where $Z$ is a standard normal random variable.  Using this expression we obtain the second moment of the pair returns:
\begin{align*}
    \mathbf E[(r_{t+t})^2] =& e^{2\mu_1+2\sigma_1^2+4\sigma^2} 
                    \paranth{\mathbf P(Z\leq -k+2\sigma)  + \mathbf P(Z\geq k+2\sigma)}\\
                    &-2e^{2\mu_1 +2\sigma_1^2+\sigma^2}\paranth{\mathbf P(Z\leq -k+\sigma)+\mathbf P(Z\geq k+\sigma)}\\
                    & + e^{2\mu_1 +2\sigma_1^2}\mathbf P(Z\geq k).
\end{align*}
The variance of the returns is then given by
\begin{align*}
    \text{Var}(r_{t+t})  = &e^{2\mu_1+2\sigma_1^2+4\sigma^2} \paranth{\mathbf P(Z\leq -k+2\sigma) + \mathbf P(Z\geq k+2\sigma)}\\
                    & -2e^{2\mu_1+2\sigma_1^2+\sigma^2}\paranth{\mathbf P(Z\leq -k+\sigma) + \mathbf P(Z\geq k+\sigma)}\\ 
                    &+2e^{2\mu_1+2\sigma_1^2}\mathbf P(Z\geq k)- \mathbf E^2[r_1].
\end{align*}

\subsection{Proof of Theorem \ref{thm:cov_cointegrated}}
We prove here the results of Theorem \ref{thm:cov_cointegrated} for the covariance of a cointegrated pair.  We consider cointegrated pairs $a$ and $b$ which share a common stock.  We assume the pair $a$ is stocks 1 and 2 and pair $b$ is stocks 1 and 3.  We can write
\begin{align*}
   r^a_{t+1}r^b_{t+1} & = S_{at}S_{bt}\paranth{e^{\mu_1 + \delta_{1,t+1}+\epsilon^{a,t+1} -\epsilon_{a,t}} -1}
   \paranth{e^{\mu_1 +\delta_{1,t+1}+\epsilon_{b,t+1} -\epsilon_{bt}} -1}\\
   &  = S^a_tS^b_t\paranth{e^{2\mu_1 +2\delta_{1,t+1} + \epsilon_{a,t+1} -\epsilon_{at} + \epsilon_{b,t+1} -\epsilon_{bt} } 
   -e^{\mu_1 +\delta_{1,t+1}+\epsilon_{a,t+1} -\epsilon_{at}}
   -e^{\mu_1 +\delta_{1,t+1}+\epsilon_{b,t+1} -\epsilon_{bt}}
   +1}.
\end{align*}
If we take the expectation of this expression  we obtain
\begin{align*}
   \mathbf E[r^a_{t+1}r^b_{t+1}] & = 
   e^{2\mu_1+2\sigma_1^2+\sigma_a^2/2+\sigma^2_b/2}\mathbf E[S_{at}e^{-\epsilon_{at}}]\mathbf E[S_{bt}e^{-\epsilon_{bt}}]\\
   & = e^{2\mu_1+2\sigma_1^2+\sigma_a^2+\sigma^2_b}
    \paranth{\mathbf P(Z\leq -k+\sigma_a) - \mathbf P(Z\geq k+\sigma_a)}
    \paranth{\mathbf P(Z\leq -k+\sigma_b) - \mathbf P(Z\geq k+\sigma_b)}\\
    & = e^{\sigma_1^2}\mathbf E[r^a_{t+1}]\mathbf E[r^b_{t+1}].
\end{align*}
Above we used the fact that the mean value of the signal for each pair is zero, equation \eqref{eq:sexp}, and Theorem \ref{thm:cointegrated_mean_var} for the mean return of the pairs.  The covariance is then given by
\begin{align*}
    \text{Cov}(r^a_{t+1},r^b_{t+1}) & = \paranth{e^{\sigma_1^2}-1}\mathbf E[r^a_{t+1}]\mathbf E[r^b_{t+1}].
\end{align*}

\subsection{Proof of Theorem \ref{thm:noncointegrated_mean_var}}
We prove here the results of Theorem \ref{thm:noncointegrated_mean_var} for the mean and variance of a non-cointegrated pair.
Consider a non-cointegrated pair of stocks 1 and 2 whose returns at time $t+1$ are given by
\begin{equation}
r_{i,t+1} = e^{\mu_i+\delta_{i,t+1}}-1, ~~i\in \curly{1,2},
\end{equation}
Note that for non-cointegrated pairs, the returns of the stocks at time $t+1$ are independent of the trading signal at time $t$.  We also note that by symmetry the signal has mean zero.  Using these properties, we have
\begin{align*}
    \E\bracket{r_{t+1}} & = \E\bracket{S_t(r_{2,t+1}-r_{1,t+1})}\\
    & = \E\bracket{S_t}\E\bracket{(r_{2,t+1}-r_{1,t+1})}\\
    & = 0.
\end{align*}
For the second moment, we have
\begin{align*}
    \mathbf E[r_{t+1}^2] & = \mathbf E[S^2_t]\mathbf E[(e^{\mu_2+\delta_{2,t+1}}- e^{\mu_1+\delta_{1,t+1}})^2]\\
    & = \mathbf E[S^2_t]\mathbf E[e^{2\mu_2+2\delta_{2,t+1}}+ e^{2\mu_1+2\delta_{1,t+1}}
    -2e^{\mu_1+\mu_2+\delta_{1,t+1}+\delta_{2,t+1}}]\\
    &=\mathbf E[S^2_t]\paranth{e^{2\mu_2+2\sigma^2_2} + e^{2\mu_1+2\sigma_1^2}
                                    -2 e^{\mu_1+\mu_2+\sigma^2_1/2+\sigma^2_2/2}}.
\end{align*}
Because of the way the signal is constructed, the expectation of it squared is
\begin{align*}
    \mathbf E [S^2_t] & = \mathbf E[ \mathbbm{1}\curly{Z \leq -k}]+\mathbf E[ \mathbbm{1}\curly{Z \geq k}]
    -2\mathbf E[ \mathbbm{1}\curly{Z \leq -k}\mathbbm{1}\curly{Z \geq k}]\\
    & = \mathbf P(Z\leq -k) + \mathbf P(Z\geq k)\\
    & = 2\mathbf P(Z\geq k)
\end{align*}
where $Z$ is a standard normal random variable.  Using this expression, we obtain the variance of the returns as
\begin{align*}
    \text{Var}(r_{t+1}) & = 2\mathbf P(Z\geq k)\paranth{e^{2\mu_1+2\sigma^2_1} + e^{2\mu_2+2\sigma^2_2}
                                    -2 e^{\mu_1+\mu_2+\sigma^2_1/2+\sigma^2_2/2}}.
\end{align*}

\subsection{Proof of Theorem \ref{thm:cov_notcointegrated}}
We prove here the results of Theorem \ref{thm:cov_notcointegrated} for the covariance of a non-cointegrated pair.  We have two non-cointegrated pairs $a$ and $b$, where pair $a$ is stocks 1 and 2 and pair $b$ is stocks 1 and 3.  Taking the expectation of the product of their returns, we obtain
\begin{align*}
   \mathbf E[r_{a,t+1}r_{b,t+1}] & = \mathbf E[S_{at}S_{bt}(e^{\mu_3+\delta_{3,t+1}}- e^{\mu_1+\delta_{1,t+1}})
                                    (e^{\mu_2+\delta_{2,t+1}}- e^{\mu_1+\delta_{1,t+1}})]\\
                                & =  \mathbf E[S^a_tS^b_t]
                                \paranth{e^{\mu_2+\mu_3 + \sigma_2^2/2 + \sigma_3^2/2}  
                                         -e^{\mu_1+\mu_3 + \sigma_1^2/2 + \sigma_3^2/2}
                                         -e^{\mu_1+\mu_2 + \sigma_1^2/2 + \sigma_2^2/2}
                                         +e^{2\mu_1 + 2\sigma_1^2}}
\end{align*}
The challenge here is with the expectation of the product of the signals.  To evaluate this expectation, we define a set of independent normal random variables $X_{it}$  with mean zero and variance $t\sigma^2_i$ for $i\in \curly{1,2,3}$.  We can express the spreads as
\begin{align*}
    \epsilon_t^a & =  X_{2t} - X_{1t}\\
    \epsilon_t^b & =  X_{3t} - X_{1t}.
\end{align*}
Let us define 
\begin{align*}
    Z_a & = \frac{X_{2t} - X_{1t}}{\sqrt{t(\sigma_1^2 + \sigma_2^2)}}\\
    Z_b & = \frac{X_{3t} - X_{1t}}{\sqrt{t(\sigma_1^2 + \sigma_3^2)}}.
\end{align*}
The random variables $Z_a$ and $Z_b$ are zero mean jointly normal random variables with covariance matrix 
\begin{align*}
    \Sigma & = \begin{bmatrix}
                1 & \frac{\sigma_1^2}{\sqrt{(\sigma_1^2 + \sigma_2^2)(\sigma_1^2+\sigma_3^2)}} \\
                 \frac{\sigma_1^2}{\sqrt{(\sigma_1^2 + \sigma_2^2)(\sigma_1^2+\sigma_3^2)}} & 1 \\
               \end{bmatrix}.
\end{align*}
Using this we can write the expected value of the signal product as
\begin{align*}
     \mathbf E[S_{at}S_{bt}]  = & \mathbf E[\paranth{\mathbbm{1}\curly{Z_a \leq -k}  }
                                        \paranth{\mathbbm{1}\curly{Z_b \leq -k}  }]
                            +  \mathbf E[\paranth{\mathbbm{1}\curly{Z_a \geq k}  }
                                        \paranth{\mathbbm{1}\curly{Z_b \geq k}  }]\\
                            &-\mathbf E[\paranth{\mathbbm{1}\curly{Z_a \leq -k}  }
                                        \paranth{\mathbbm{1}\curly{Z_b \geq k}  }]
                            -  \mathbf E[\paranth{\mathbbm{1}\curly{Z_a \geq k}  }
                                        \paranth{\mathbbm{1}\curly{Z_b \leq -k}  }]\\
                        =& 2\mathbf P\paranth{Z_a\geq k \bigcap Z_b\geq k }
                          -2\mathbf P\paranth{Z_a\geq k \bigcap Z_b\leq -k }.
\end{align*}
Noting that the mean value of the returns of a non-cointegrated pair is zero  from Theorem \ref{thm:noncointegrated_mean_var}, we can then write the covariance as
\begin{align*}
    \text{Cov}(r_{a,t+1}, r_{b,t+1}) & = \mathbf E [r_{a,t+1}r_{b,t+1}]\\
                     =&  2\paranth{\mathbf P\paranth{Z_a\geq k \bigcap Z_b\geq k }
                          -\mathbf P\paranth{Z_a\geq k \bigcap Z_b\leq -k }}\\
                          &\times\paranth{e^{\mu_2+\mu_3 + \sigma_2^2/2 + \sigma_3^2/2}  
                                         -e^{\mu_1+\mu_3 + \sigma_1^2/2 + \sigma_3^2/2}
                                         -e^{\mu_1+\mu_2 + \sigma_1^2/2 + \sigma_2^2/2}
                                         +e^{2\mu_1 + 2\sigma_1^2}}.
\end{align*}

\end{document}